\documentclass{pasa}%

\title[MWA EoR1 foregrounds]{A high resolution foreground model for the MWA EoR1 field: model and implications for EoR power spectrum analysis}
\author[Procopio et al.]{P.~Procopio$^{1,2}$, R.~B.~Wayth$^{2,3}$, J.~Line$^{1,2}$, C.~M.~Trott$^{2,3}$, H.~T.~Intema$^{4,6}$, D.~A.~Mitchell$^{7}$,B.~Pindor$^{1,2}$, J.~Riding$^{1}$, S.~J.~Tingay$^{3,5}$, M.~E.~Bell$^{2,7}$, J.~R.~Callingham$^{14}$, K.~S.~Dwarakanath$^{9}$, Bi-Qing For$^{10}$, 
B.~M.~Gaensler$^{2,8,11}$, P.~J.~Hancock$^{3}$, L.~Hindson$^{12}$, N.~Hurley-Walker$^{3}$, M.~Johnston-Hollitt$^{12,13}$, A.~D.~Kapi\'{n}ska$^{2,10}$, 
 E.~Lenc$^{2,8}$, B.~McKinley$^{1,2}$, J.~Morgan$^{3}$, A.~Offringa$^{14}$, L.~Staveley-Smith$^{2,10}$, Chen~Wu$^{10}$, \and Q.~Zheng$^{12,13}$\\
\affil{$^{1}$School of Physics, The University of Melbourne, Parkville, VIC 3010, Australia}
\affil{$^2$ARC Centre of Excellence for All-Sky Astrophysics (CAASTRO)}
\affil{$^3$International Centre for Radio Astronomy Research (ICRAR), Curtin University, Bentley, WA 6102, Australia}
\affil{$^{4}$National Radio Astronomy Observatory, Socorro, NM, USA}
\affil{$^{5}$INAF, Istituto di Radioastronomia, Via Piero Gobetti, I-40129 Bologna, Italy}
\affil{$^{6}$Leiden University, The Netherlands}
\affil{$^{7}$CSIRO Astronomy and Space Science (CASS), PO Box 76, Epping, NSW 1710, Australia}
\affil{$^{8}$Sydney Institute for Astronomy, School of Physics, The University of Sydney, NSW 2006, Australia}
\affil{$^{9}$Raman Research Institute, Bangalore 560080, India}
\affil{$^{10}$International Centre for Radio Astronomy Research, University of Western Australia, Crawley, WA 6009, Australia}
\affil{$^{11}$Dunlap Institute for Astronomy and Astrophysics, University of Toronto, 50 St.\ George Street, Toronto, ON M5S 3H4, Canada}
\affil{$^{12}$School of Chemical \& Physical Sciences,  Victoria University of Wellington, Wellington 6140, New Zealand}
\affil{$^{13}$Peripety Scientific Ltd., PO Box 11355 Manners Street, Wellington 6142, New Zealand}
\affil{$^{14}$ASTRON, The Netherlands Institute for Radio Astronomy, Postbus 2, 7990 AA, Dwingeloo, The Netherlands}
}
\jid{PASA}
\doi{10.1017/pas.\the\year.xxx}
\jyear{\the\year}

\usepackage[authoryear]{natbib}
\bibpunct{(}{)}{;}{a}{}{,}
\usepackage{array,graphicx,color}
\usepackage{url}
\usepackage{subcaption}

\begin{document}

\begin{abstract}

The current generation of experiments aiming to detect the neutral hydrogen signal from the Epoch of Reionisation (EoR) is likely to be limited by systematic effects associated with removing foreground sources from target fields. In this paper we develop a model for the compact foreground sources in one of the target fields of the MWA's EoR key science experiment: the `EoR1' field. The model is based on both the MWA's GLEAM survey and GMRT 150\,MHz data from the TGSS survey, the latter providing higher angular resolution and better astrometric accuracy for compact sources than is available from the MWA alone.
The model contains 5049 sources, some of which have complicated morphology in MWA data, Fornax A being the most complex.
The higher resolution data show that 13\% of sources that appear point-like to the MWA have complicated morphology such as double and quad structure, with a typical separation of 33~arcsec.
We derive an analytic expression for the error introduced into the EoR two-dimensional power spectrum due to peeling close double sources as single point sources and show that for the measured source properties, the error in the power spectrum is confined to high $k_\bot$ modes that do not affect the overall result for the large-scale cosmological signal of interest. The brightest ten mis-modelled sources in the field contribute 90\% of the power bias in the data, suggesting that 
it is most critical to improve the models of the brightest sources.
With this hybrid model we reprocess data from the EoR1 field and show a maximum of 8\% improved calibration accuracy and a factor of two reduction in residual power in $k$-space from peeling these sources.
Implications for future EoR experiments including the SKA are discussed in relation to the improvements obtained.

\end{abstract}
\begin{keywords}
Reionization -- techniques: interferometric -- radio continuum: galaxies -- large-scale structure of Universe
\end{keywords}
\maketitle%
\section{INTRODUCTION }
\label{sec:intro}

A key science goal for current and next-generation low-frequency radio telescopes is to make a measurement of the power spectrum of the faint radio signals from neutral hydrogen in the early universe \citep{parsons10,2013PASA...30...31B,vanhaarlem13,koopmans15,deboer16}.
These experiments are very challenging in several ways. In addition to basic signal-to-noise requirements demanding thousands of hours of observing time, the experiments must deal with the so-called ``foregrounds'' (which are essentially all other sources of radio emission) that can potentially eliminate a power spectrum detection or bias a measurement.
Strategies to understand and correct for the effects of foregrounds, including how foregrounds interact with the instrument response, are the subject of many investigations \citep{morales2006,jelic2008,jelic10,2009A&A...500..965B,chapman12,trott12,morales12,vedantham12,thyagarajan15a,ewallwice2016,2016MNRAS.461.3135B,2016PASA...33...19T,2016MNRAS.463.4317P}.
Nevertheless, the bright foreground sources (such as diffuse Galactic radio emission and extragalactic radio galaxies and quasars) are spectrally smooth \citep{dimatteo2002,2014PhRvD..89b3002D,2014ApJ...788..106P,zalda2004,2016MNRAS.458.1057O} which, in principle, confines the signal from these sources to the lowest-order (slowest varying) terms in the 1/frequency dimension of a power spectrum.

The Murchison Widefield Array \citep[MWA,][]{2013PASA...30....7T} was designed with an EoR power spectrum detection as one of the key science goals \citep{2013PASA...30...31B}. 
The MWA's EoR science team is taking a multi-pronged approach to the experiment, pursuing the detection of the cosmological signal through the development of multiple independent pipelines, characterised by different approaches and techniques \citep{jacobs16}. A detailed overview of the of one of these approaches can be found in \citet{2016ApJ...833..102B}, which also contains results obtained on a complementary MWA EoR field.

One of the ideas explored in \citet{jacobs16} exploits detailed knowledge of foregrounds, constituting the driving idea behind the work carried out in this paper. In particular, we want to evaluate if the inclusion of extra, higher-resolution data helps in better modelling and subtracting the foregrounds and explore the benefits for the pipeline considered in terms of output data quality.

Among the compact foreground sources included in the sky model created in this work, approximately $13\%$ are partially resolved at MWA resolution and/or consist of multiple unresolved components at higher angular resolution. This fraction includes the brightest sources of the catalogue, which affect the calibration process more heavily. Using a single component model for these sources leaves residuals after subtraction, which translates to residual structures in image and Fourier space, contaminating the underlying EoR signal. 


The Real Time System \citep[RTS,][]{2008ISTSP...2..707M} and the Cosmological H I Power Spectrum Estimator \citep[CHIPS,][]{trott16} constitute one of the two pipelines used in the MWA EoR power spectrum experiment. Calibration and source subtraction are performed by the RTS while the power spectrum (PS) estimation is carried out by CHIPS. Aiming to improve the first half of the pipeline, the RTS was implemented to properly ingest and handle models with multiple source components for a more accurate representation of the sky model. The input catalogue can be composed of a mixture of different entries, such as multi-component models (point source or Gaussian) or shapelet models. This RTS feature is of key importance in our analysis, allowing for a straightforward inclusion of the higher resolution data. 

In this paper we present a high-resolution foreground model for the MWA EoR1 field (centred at RA=$4^h$, Dec=$-30^d$) and discuss the results obtained with it and the implications for the EoR PS experiment. In Sec \ref{sec:data} we present the data used and show some details of the processing. In Sec \ref{sec:mis-sub} a prediction on how source mis-subtraction affects the PS analysis is performed and the results compared with real data. The effects of the improved model on calibration and source subtraction are discussed in Sec. \ref{sec:imp_mod}, along with more details from the PS analysis. Finally, in Sec \ref{sec:disc} and \ref{sec:conc} we summarise the results obtained and discuss their implications for other experiments devoted to EoR signal detection.

\section{A combined high-resolution foreground model}
\label{sec:data}
\subsection{MWA data}
\label{sec:mwadata}

We used data from the 2013 October observing campaign of the `EoR1' field.
This field was observed in two frequency bands: the low and the high band, ranging from $138.9$ to $169.6$ MHz and from $167.0$ to $197.7$ MHz respectively. Only high band data were used in this paper (see Table~\ref{table:observations} for further details). The selected band guarantees the best available angular resolution for MWA EoR observation, $2^{\prime}$. 
We avoided nights with higher than usual ionospheric activity.
In total, 191 observations of length 112\,s were included in the subset, totalling $\sim6$\,h of data. 

These observations constitute the dataset which the different iterations of the sky model are tested on. 

\newcolumntype{P}[1]{>{\centering\arraybackslash}p{#1}}
\begin{table}
\renewcommand{\arraystretch}{1.2}
\centering
\caption{Details of the MWA data used. Only the central frequency of the observing band is here reported.}
	\begin{tabular}{P{2cm} P{2cm}P{3cm}}
	\hline
	\textbf{Night} & \textbf{Frequency} & \textbf{GPS time range} \\
	\hline \hline
	2 Oct 2013 & 181 MHz & 1064771280 - 1064779824 \\
	8 Oct 2013 & 181 MHz & 1065289488 - 1065297656 \\
	23 Oct 2013 & 181 MHz & 1066580664 - 1066592984 \\
	\hline \hline
	\end{tabular}
\label{table:observations}
\end{table}

\subsection{TGSS data}
\label{sec:TGSSdata}

The TIFR GMRT Sky Survey\footnote{\url{http://tgss.ncra.tifr.res.in}} (TGSS) Alternative Data Release 1 \citep[ADR1,][]{intema17} represents an independent re-processing of the $150$ MHz continuum survey from the Giant Metrewave Radio Telescope \citep[`GMRT',][]{1991ASPC...19..376S} that was carried out between 2010 and early 2012.
Although the mosaic used for this work was composed before the official TGSS release, the crucial steps of the data processing pipeline are those described in \citet{intema17}.
Here we limit ourselves to giving a brief summary of the process.

The core of the pipeline is represented by the Source Peeling and Atmospheric Modeling (SPAM) package \citep{2009A&A...501.1185I}, which makes use of direction-dependent calibration, modelling and imaging, primarily to correct for the dispersive delay introduced by ionospheric activity. 

During the pre-processing phase, excessive radio frequency interference (RFI) is flagged, followed by estimation of gain and bandpass calibration parameters. This process is repeated several times with increasingly tight RFI flagging thresholds to improve the initial calibration solution. In the main pipeline further steps of calibration, flagging, and wide-field imaging are performed to produce the final images. The direction-independent gain calibration (phase only) is carried out on a 16\,s timescale, tracking in this way the time-varying effects of the ionospheric phase delay. Wide-field imaging for self-calibration purposes (including further RFI flagging) is then performed, using Briggs weighting with the robust parameter set to $-1$. It should be noted that short baselines were also flagged, hence emission will not extend beyond $10^\prime-20^\prime$ in the final images.
Finally, direction-dependent gain phases are obtained, characterising multiple single sources against the ionospheric phase delay. During this phase, an ionosphere model is fitted to the data per time interval. This is used to generate gain tables for each of the small facets covering the primary beam, which are then used to perform the ionospheric corrections per antenna per time stamp. 

The final images of each pointing are combined into $5^{\circ}\times5^{\circ}$ mosaics for further processing. Maintaining high spatial resolution and well-behaved global properties of the restoring beam are of crucial importance. Due to the changing shape of the restoring beam associated with the pointing DEC, two different mosaic beams were adopted; a circular beam for the sky north of the GMRT latitude and a N-S elongated beam for pointings south of the GMRT latitude. For each mosaic, its beam is fully defined by the mosaic centre DEC. Overlapping images are convolved and renormalised to match the resolution and orientation of the mosaic beam. 

The final mosaic is $11612 \times 11612$ pixels with pixel scale $6.2^{\prime\prime}$, corresponding roughly to a $20^{\circ}\times20^{\circ}$ region centred on the EoR1 field.

\subsection{Source extraction and cross-matching}
\label{sec:src_extraction}

We used source positions from the extragalactic compact source catalogue \citep{Hurley-Walker2016} from the GaLactic and Extragalactic All-Sky MWA (GLEAM) survey \citep{2015PASA...32...25W} as the base for building our sky model. The GLEAM survey frequency coverage spans from $72$ to $231\,$MHz, hence it overlaps the TGSS data used for this work.
We initially treat the sky model as if it is composed by point sources only, with the exception of Fornax A, which is an extremely bright ($> 100\,$Jy at relevant frequencies) source with a complex morphology. As a point source model is insufficient for calibration/subtraction, a shapelet model for Fornax A was employed from the beginning and remains unchanged for every catalogue iteration and version discussed in this paper. More details on the model are given in Sec. \ref{sec:fnxa}.

We aim to improve the models of the extended and partially resolved sources located in the MWA EoR1 field by using the higher resolution TGSS data. As a starting point, a blind source extraction was performed on the TGSS mosaic using PyBDSM\footnote{\url{http://www.astron.nl/citt/pybdsm/}} to create a list of sources to be cross-matched with the base catalogue. Investigating the outcomes of this cross-match enabled us to identify any potentially extended or partially resolved sources that appear as a single component in the base GLEAM catalogue, due to the lower resolution of the MWA. During this source extraction, we left the PyBDSM parameters in their default value, with the only exception of using an adaptive RMS box due to slightly changing background features in the TGSS mosaic. 

To build a sky model with reliable spectral information, along with the TGSS catalogue generated in this work, the GLEAM catalogue was cross-matched to the following catalogues: the $74\,$MHz Very Large Array Low Frequency Sky Survey redux~\citep[VLSSr,][]{Lane2012}; the $408\,$MHz Molonglo Reference Catalogue~\citep[MRC,][]{Large1981}; the $843\,$MHz Sydney University Molonglo Sky Survey~\citep[SUMSS,][]{Mauch2003}; and the $1.4\,$GHz NRAO VLA Sky Survey~\citep[NVSS,][]{Condon1998}. 

The cross-matching was performed using the Positional Update and Matching Algorithm ~\citep[PUMA,][]{Line2017}. PUMA is specifically designed to cross-match low radio-frequency ($\leq\sim1\,$GHz) catalogues by using both positional and spectral data. As it is also designed to cross-match catalogues generated with surveying instruments with different resolutions, it is well-placed to identify partially and fully-resolved sources in MWA data, by correctly cross-matching multiple sources from higher resolution catalogues.

To identify possible cross-matches between the catalogues, GLEAM was initially cross-matched to all other catalogues using an angular cut-off of $2.5$ arcmin (approximately the FWHM of the MWA synthesized beam). PUMA then assesses the positional probability of a match, as well as investigating the resultant spectral energy distribution, and assigns the following matching classifications to each GLEAM source \citep[see][for details]{Line2017}:
\begin{list}{•}{•}
\item[\texttt{isolated}] - the source is unresolved in all catalogues, or has no nearby confusing sources; a straight forward cross-match.
\item[\texttt{dominant}] - there are multiple possible cross-match candidates from one or more of the cross-matched catalogues; this is a confused cross-match. Based on fitting to a power-law model, there is one particular cross-match (involving one catalogued entry from each catalogue) that well describes the source.
\item[\texttt{multiple}] - if there is no \texttt{dominant} cross-match, the GLEAM catalogue is likely blending multiple sources that are resolved in the other catalogues. To test if all cross-matches are from the same astrophysical source (for example a double-lobed radio galaxy), the flux densities of all matched sources from the same matched catalogue are summed. If this combined spectral information fits a power-law well, accept this combined cross-match.
\item[\texttt{eyeball}] - if all matching criteria fail, the source likely has a complex or extended morphology, and the match is flagged for visual inspection.
\end{list}


\subsection{Cross-match results}
\label{subsec:crossmatch-results}
\begin{figure}
\includegraphics[width=\columnwidth]{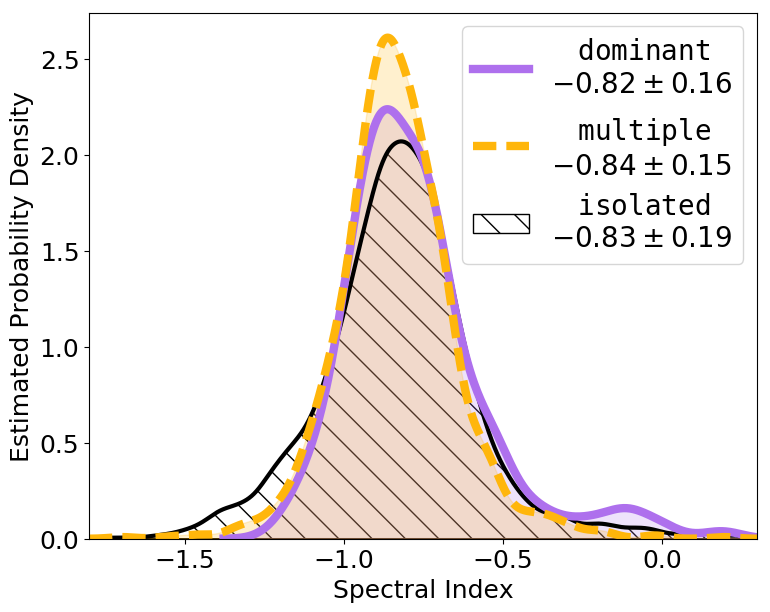}
\caption{A KDE for each PUMA matching classification. The KDE technique uses a smoothing kernel to non-parametrically estimate the probability density function of a random variable. As the width of the smoothing function is estimated from the data (see text in Sec~\ref{subsec:crossmatch-results}), statistically significant trends in the data should be highlighted. These can be surpressed in a histogram due to the discontinous nature of the binning involved. The legend includes the median and median absolute deviation for each distribution.}
\label{fig:SIdist}
\end{figure}

7598 GLEAM sources were matched within the TGSS field, of which: 81\% were \texttt{isolated}; 2\% were \texttt{dominant}; 16\% were \texttt{multiple};  1\% were \texttt{eyeball}. To check that the automatically matched PUMA outcomes were consistent and sensible, a power-law was fitted to every spectrum\footnote{We assume that the underlying astrophysical processes are similar in nature for the considered source.} and the spectral index (SI) distribution of each matching type was investigated as shown in Figure~\ref{fig:SIdist}. The kernel density estimates (KDEs) were generated using a univariate estimator with a Gaussian kernel with a bandwidth calculated from the data using Scott's rule of thumb~\citep{Scott1992}. We find a consistent median and similar distributions for each matching type.


\subsection{Properties of GLEAM compact sources with matches to TGSS sources}
\label{sec:properties_of_GLEAM_sources}
\begin{figure}
\includegraphics[width=\columnwidth]{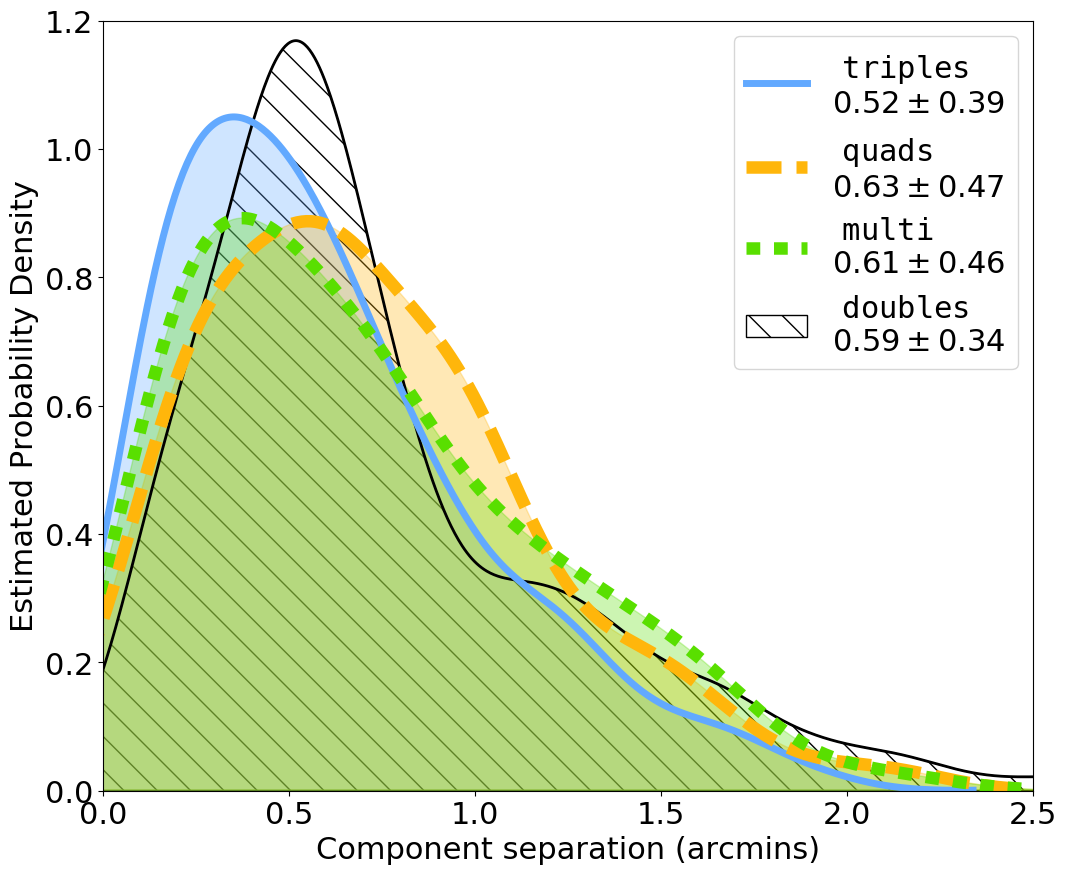}
\caption{KDEs of the separation of multiple TGSS sources matched to a single GLEAM source, grouped as detailed in Table~\ref{table:num_comps}. The legend includes the median and median absolute deviation for each distribution. As each plotted distribution is a non-parametric estimate made from the data, the combination of the Gaussian kernel and bandwidth allows the derived distribution to extended to negative separations. These are, of course, non-physical but allow the density estimate to fall to zero without using prior constraints on the fit.}
\label{fig:GMRTseps}
\end{figure}

As the TGSS frequency lies within the GLEAM band, the two catalogues lend themselves to a direct morphological comparison. A total of 5049 of the 7598 GLEAM sources were matched to sources found by PyBDSM in the TGSS mosaic; Table~\ref{table:num_comps} shows the number of TGSS sources matched to each single GLEAM source. Figure~\ref{fig:GMRTseps} shows the angular separation of TGSS sources matched to a single GLEAM source, showing a strong peak separation distance of $\sim30''$ regardless of the number of matched TGSS sources. This value is consistent with the nominal 25" angular resolution of TGSS, and we would not expect sources to be identified as doubles below this resolution. Physically, the distribution of doubles would extend to zero separation.

Table~\ref{table:num_comps} shows that the majority of GLEAM sources are point-like at the resolution of TGSS ($\sim87$\%), however the majority of the sources that have multiple TGSS components are doubles ($\sim11$\%). Motivated by this, as well as investigating the effects of adding in extended models, we also investigate the effects of introducing models for close double sources. In \S\ref{sec:mis-sub}, we analytically make a prediction of this effect with which to compare our results.

\newcolumntype{P}[1]{>{\centering\arraybackslash}p{#1}}
\begin{table}
\renewcommand{\arraystretch}{1.2}
\centering
\caption{The number of TGSS sources matched to a single GLEAM source.}
	\begin{tabular}{P{3cm} P{2cm}}
	\hline
	\textbf{Number of TGSS sources} & \textbf{Number of instances}  \\
	\hline \hline
	1 & 4368 \\
	2 & 545  \\
	3 & 91 \\
	4 & 26  \\
	$>$4 & 19 \\
	\hline \hline
	\end{tabular}
\label{table:num_comps}
\end{table}

\subsection{Assembling extended source models}

\begin{figure}
\includegraphics[width=\columnwidth]{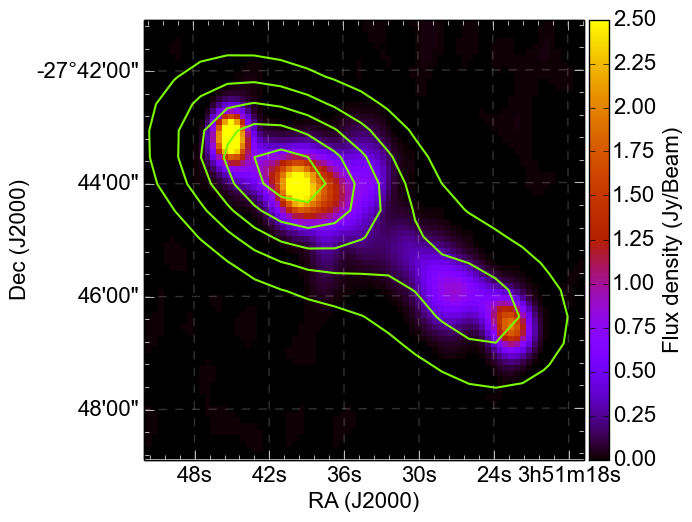}
\caption{The source PMN J0351-2744 as it appears in the TGSS ADR1 data, with a contour plot of MWA EoR1 data overlaid.}
\label{fig:res_comp}
\end{figure}

Using the results of the cross-matches found in \S\ref{subsec:crossmatch-results}, we created an extended source model for any source classified as \texttt{eyeball} by PUMA. To create these models, a further processing of the TGSS mosaic though PyBDSM was required. Due to the different morphologies of the sources considered, more human interaction was required through this processing phase. Using the source coordinates we first trimmed a box of a few arc-minutes (actual value depending of the angular size of the source) around each target. Then we make use of some PyBDSM parameters for a fitting process more inclined to extended sources. In particular, we activate the wavelet decomposition on multiple scales of the Gaussian residuals,  we increase the area value a Gaussian needs to have to be flagged, and we set the output format to Gaussian, so each component of each outputted source is characterised by its own major/minor axis and PA. Regardless of their angular size, all of the sources modelled through PyBDSM were treated as Gaussians or clusters of Gaussians. Figure~\ref{fig:res_comp} shows the dramatic difference in resolution between the two datasets used in this work.

Three separate sky models were produced to test the effects of introducing extended high resolution models as well as attempting to verify the analytic results derived in \S\ref{sec:mis-sub}. In detail, these were:
\begin{list}{•}{•}
\item[\textbf{Point source model}] - a single point source model was included for every GLEAM source, including the complex sources classified as \texttt{eyeball} by PUMA. For these 114 sources, a point source model was created by hand using catalogued information. The spectrum of each source was fitted with a 2$^\mathrm{nd}$-order polynomial using weighted least squares to ensure smooth spectral behaviour in the RTS.
\item[\textbf{Split double models}] - every GLEAM source matching two TGSS sources was split into a double point source, based on the TGSS positions and fluxes. The flux densities at other frequencies were divided between the two new components by weighting by the TGSS flux densities. This automatically gives precisely the same spectral behaviour for both components which may not be physically the case, but by weighting by a flux density at the frequencies at which peeling is occurring, the impact is considered minimal.
\item[\textbf{Extended source model}] - for the 114 complex sources, a multiple Gaussian model was created based on the TGSS data. All other sources in the model were as in the split doubles model for consistency.
\end{list}

\subsection{Fornax A}
\label{sec:fnxa}

The EoR1 field hosts one of the brightest radio sources in the Southern sky: Fornax A. It is far more extended than any other source in the field, and has the largest flux density. The need for an accurate model for such a complex source motivated the use of the RTS with its capability to ingest a sky model composed of a mixture of source model types. In particular, it has been designed to take shapelet models for extended sources in order to ensure a robust handling of the source morphology during calibration and peeling.
  
A separate analysis was carried out to obtain a model for the radio galaxy Fornax A. We used a subset of the MWA data to create a high-resolution CLEANed image of the region around the source, and processed it through a shapelet decomposition code.
The recovered model was then included in all the sky models used during the processing to avoid any discrepancy that could be introduced using different models for such a strong source.



We note that the TGSS imaging parameters were tuned in such way to optimise the shape of the point spread function (PSF). This, in addition to the flagging of the short baselines, came at the cost of losing sensitivity to extended sources. Hence the large radio lobes of Fornax A are resolved out in the TGSS mosaic, so no additional information from the TGSS images is available.


\section{Impact of mis-subtracting close doubles as point sources}
\label{sec:mis-sub}

We aim to understand the effect of subtracting a population of closely-spaced doubles as point source on the two-dimensional (cylindrically-averaged) EoR power spectrum of brightness temperature fluctuations. The cylindrically-averaged power spectrum computes the variance in the signal as a function of angular ($k_\bot$) and line-of-sight ($k_\parallel$) spatial wavemodes. Although we expect the 21~cm signal to be isotropic (up to velocity-space distortions), separating these perpendicular components allows observers to discriminate spectrally-smooth foreground contaminants, such as continuum sources, from spectrally-structured 21~cm fluctuations.

Motivated by the distribution of source components found in section \ref{sec:properties_of_GLEAM_sources}, we model the underlying angular separation of multi-component GLEAM sources by a Rayleigh distribution.
The Rayleigh distribution has desirable and physically motivated features, including positive-only separations, and a Gaussian-like peak with an extended large separation tail.
Its probability distribution function is described by:
\begin{equation}
\mathcal{R}(\theta;\sigma) = \frac{\theta}{\sigma^2}e^{\left(-\frac{\theta^2}{2\sigma^2}\right)},
\end{equation}
where $\sigma$ characterises the distribution and the mode and variance are given, respectively, as $\mu=\sigma$ and var$=(4-\pi)/2\sigma^2$.
The data-estimated distribution of double sources from Figure \ref{fig:GMRTseps} is well-fitted with $\sigma=33$~arcsec (0.55~arcmin), and we assume that the measured fraction of GLEAM sources that are actually close doubles $\xi\equiv$~545/5049~=~11\% is representative of the full sky.

To estimate the statistical signature of mis-subtracted doubles, we build from the existing framework of \citet{trott16}, which takes a spatially Poisson-distributed population of spectrally-smooth sources in the sky, and computes their power in the power spectrum, considering a full, frequency-dependent instrument model (baseline sampling and primary beam). This framework yields the familiar `wedge'-like structure in the power spectrum parameter space, whereby the low $k_\bot$ modes are contaminated only in the DC ($k_\parallel=0$) mode, and this contamination extends into higher $k_\parallel$ modes for larger $k_\bot$ modes, i.e., smaller angular scales \citep{datta10,trott12,thyagarajan13,vedantham12}. To extend this model for mis-subtracted point sources, we follow the following procedure:
\begin{enumerate}
\item Compute the residual visibility that is produced from subtracting a single, centred point source visibility from the actual visibility formed from two closely-spaced sources, each with half of the point source flux density;
\item Describe the variance of a visibility from doubles contained within a small differential region of the sky, where the doubles have random orientations, Rayleigh-distributed separations, and a flux density distribution that matches that for measured point sources;
\item Compute the full variance from all doubles in that differential region by integrating over the orientation on the sky, and separation of the doubles, and multiplying by the fraction of apparent point sources that are actually doubles;
\item Compute the covariance between visibilities measured in different frequency channels, by integrating over the primary beam-weighted field-of-view and flux density distribution.
\end{enumerate}
At the final step we obtain the data covariance matrix for the residual signal from mis-subtracting doubles as point sources. We highlight the main equations here, and provide the full derivation in the Appendix.

For a sky with only one double-source, with a centre-of-mass location of $(l,m)$ with total flux density, $S$, the visibility at wavenumber $(u,v)$ wavelengths is given by
\begin{eqnarray}
&V&(u,v)\\ \nonumber
&=& \frac{S}{2}e^{(-2\pi{i}(u(l\!-\!\Delta{l}/2)\!+\!v(m\!-\!\Delta{m}/2))} \\\nonumber
&& + \frac{S}{2}e^{(-2\pi{i}(u(l\!+\!\Delta{l}/2)\!+\!v(m\!+\!\Delta{m}/2))} \\
&=& Se^{(-2\pi{i}(ul+vm))}\\\nonumber
&& \times \left( e^{\pi{i}(u\Delta{l}+v\Delta{m})} + e^{-\pi{i}(u\Delta{l}+v\Delta{m})} \right)\\
&=& V_{\rm point}(u,v)\left( \cos{\pi(u\Delta{l} + v\Delta{m})} \right),
\end{eqnarray}\normalsize
where `point' indicates the visibility for a point source at that location, and $\Delta{l}, \Delta{m}$ denote the source separation ($\Delta{r}=\sqrt{\Delta{l}^2+\Delta{m}^2}$). Here, the summation of the two complex exponentials has been reduced to its cosine form. The residual visibility from mis-subtraction is therefore the difference between this expression and that for the point source, yielding;
\begin{equation}
V_{\rm res}(u,v) = V_{\rm point}(u,v)\left( \cos{\pi(u\Delta{l} + v\Delta{m})}-1 \right).
\end{equation}
This residual visibility can be extended to include a distribution of point sources co-located in a small differential sky area, with a flux density distribution given by an empirical power law\footnote{$\frac{dN}{dS}=\alpha S^{-\beta}$~/Jy/sr \citep{intema11}.}. The variance on this visibility in different patches of sky can be formed by recalling that the variance of a Poisson-distributed variable is its mean.
For doubles located in differential sky area, $(l+dl,m+dm)$, this variance on a visibility due to the Poisson-distribution of randomly-oriented and Rayleigh-separated doubles is given by:
\begin{eqnarray}
&{\rm var}&(V(u,v)) = \int_{S_{\rm min}}^{S_{\rm max}} S^2\frac{dN}{dS}dS \, \int_{2\pi} p(\phi)d\phi \\\nonumber 
&& \int_{\Delta{r}} p(\Delta{r})d\Delta{r} \, (\cos{\Delta{r}\pi(u\cos{\phi} + v\sin{\phi})}-1)^2\\\label{eqn:probs}
&=& \frac{\alpha}{3-\beta}\frac{S_{\rm max}^{3-\beta}}{S_0^{-\beta}} \, \int_{2\pi} p(\phi)d\phi  \\\nonumber
&& \int_{\Delta{r}} p(\Delta{r})d\Delta{r} \, (\cos{\Delta{r}\pi(u\cos{\phi} + v\sin{\phi})}-1)^2,
\end{eqnarray}
where,
\begin{eqnarray}
\phi &\sim& \mathcal{U}(0,2\pi] = p(\phi)\\
\Delta{r} &\sim& \mathcal{R}(r;\sigma) = p(\Delta{r}),
\end{eqnarray}
are the uniform and Rayleigh distribution, and $\alpha=4100\,{\rm Jy}^{-1}{\rm sr}^{-1}$ and $\beta=1.59$ characterise the power-law flux density distribution \citep{intema11,gervasi08}. We linearise the squared-cosine term for small separations, and integrate to find (see Appendix):
\begin{equation}
{\rm var}(V_{\rm res}(u,v)) = \frac{\alpha}{3-\beta}\frac{S_{\rm max}^{3-\beta}}{S_0^{-\beta}} \,\frac{3\pi^5}{8}(u^2+v^2)^2\sigma^4.
\end{equation}
As intuitively expected, the error term grows for larger baselines and separations, in line with the expectations for the sampling of small-scale structures on the sky.

This expression describes the additional variance for a given measurement due to a distribution of close-spaced doubles located within a differential sky area, which have been subtracted as point sources. For the power spectrum, one Fourier Transforms the spectral (line-of-sight) measurements to obtain the line-of-sight wavenumber, $\eta\propto{k}_\parallel$. To perform this step, one needs the spectral covariance (frequency-frequency covariance) of each $(u,v)$ sample, in order to correctly propagate the correlations between frequency channels. To determine the covariance within the primary field-of-view of the instrument (and attenuated by its sky response), we extend to (see Appendix for full derivation):
\begin{eqnarray}
&&{\bf C}_{\rm res}(u,v;\nu,\nu^\prime) = \frac{\alpha}{3-\beta}\frac{S_{\rm max}^{3-\beta}-S_{\rm min}^{3-\beta}}{S_0^{-\beta}}\left( \frac{3\pi^5}{8}(u^2+v^2)^2\sigma^4 \right) \\\nonumber
&& \times \displaystyle\iint d{\bf l} \, B({\bf l};\nu)B({\bf l};\nu^\prime) \, e^{\left( \frac{-2{\pi}i}{\nu_0}\Delta\nu({\bf u}\cdot{\bf l}) \right)}.
\end{eqnarray}
For a frequency-dependent, Gaussian-shaped beam, with characteristic width, $A(\nu)=(c\epsilon)/(\nu{D})$, with $\epsilon\simeq{0.42}$, the contribution of a $\xi$ fraction of closely-spaced doubles, with flux densities in the range, $[{\rm S_{min}},{\rm S_{max}}]$, to the foreground covariance is given by:
\begin{eqnarray}
&&{\bf C}_{\rm res}(u,v;\nu,\nu^\prime) = \frac{\alpha\xi}{3-\beta}\frac{(S_{\rm max}^{3-\beta}-S_{\rm min}^{3-\beta})}{S_0^{-\beta}}\frac{\pi{c^2}\epsilon^2}{D^2}\label{cov_nunu}\\\nonumber
&&\times \left(3\pi^5(u^2+v^2)^2\sigma^4 \right) \frac{1}{8(\nu^2 + \nu^{\prime{2}})} e^{\left( \frac{-u^2c^2f(\nu)^2\epsilon^2}{4(\nu^2 + \nu^{\prime{2}})D^2} \right)}.
\end{eqnarray}
In a similar vein, we can define the regular, point-source only covariance matrix:
\begin{eqnarray}
{\bf C}_{\rm PNT}(u,v;\nu,\nu^\prime) &=& \frac{\alpha}{3-\beta}\frac{S_{\rm max}^{3-\beta}}{S_0^{-\beta}} \frac{\pi{c^2}\epsilon^2}{D^2} \label{cov_nunupnt}\\\nonumber
&\times& \frac{1}{\nu^2 + \nu^{\prime{2}}} e^{\left( \frac{-u^2c^2f(\nu)^2\epsilon^2}{4(\nu^2 + \nu^{\prime{2}})D^2} \right)}.
\end{eqnarray}
The contribution to the power spectrum is then equation \ref{cov_nunu} propagated through the spectral Fourier Transform, $\mathcal{F}$:
\begin{equation}
P_{\rm res}(u,v,\eta) = {\rm diag} \left(\mathcal{F}^\dagger {\bf C}_{\rm res}(u,v;\nu,\nu^\prime) \mathcal{F} \right),
\label{eqn:fourier_prop}
\end{equation}
where we define the Fourier convention as:
\begin{equation}
\mathcal{F}(f(x)) = \tilde{f}(l) = \Delta\nu \, \displaystyle\sum_{k=0}^{N-1} f(x_k)\,e^{\left(\frac{-2\pi{i}k}{N-1} \right)},
\label{eqn:fourier}
\end{equation}
and $N$ is the number of channels with $\Delta\nu$ spectral resolution per channel. Finally, having computed this expression, we cylindrically-average $u$ and $v$ modes to yield ($k_\perp^2=u^2+v^2$), and bin into the 2D power spectrum.

\begin{figure*}[t]
{
\includegraphics[width=0.9\textwidth,angle=0]{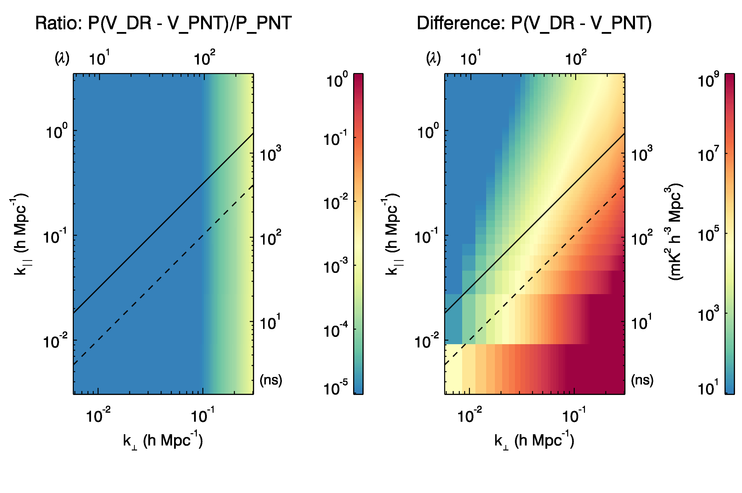}
}
\caption{(Left) \textit{Predicted} ratio of residual power in the power spectrum ($P_{\rm res}=P(V_{\rm DR} - V_{\rm PNT})$) when closely-spaced doubles are subtracted as double sources, relative to when they are subtracted as point sources ($P_{\rm PNT}$); (Right) Power in residual visibilities when peeling non-point sources correctly, and as point sources ($P(V_{\rm DR} - V_{\rm PNT})$). }
\label{fig:residual_power}
\end{figure*}

\begin{figure*}[t]
{
\includegraphics[width=0.9\textwidth,angle=0]{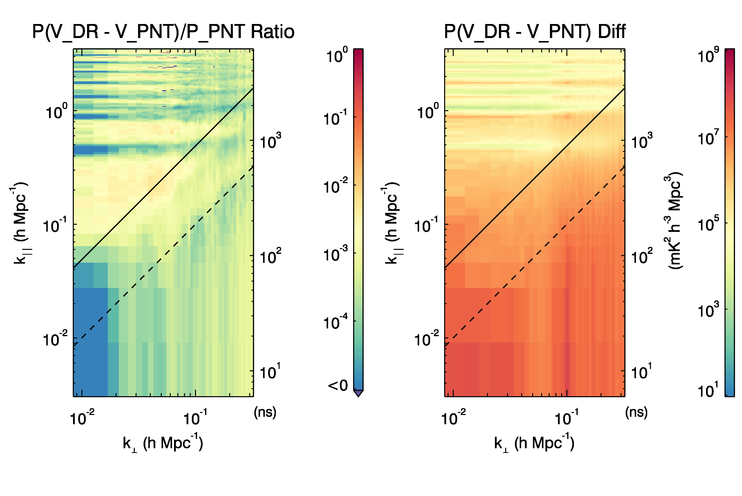}
}
\caption{(Left) \textit{Data}: Ratio of residual power in the power spectrum when closely-spaced doubles are subtracted as double sources, relative to when they are subtracted as point sources; (Right) Power in residual visibilities when peeling non-point sources correctly, and as point sources ($P(V_{\rm DR} - V_{\rm PNT})$).}
\label{fig:residual_power_data}
\end{figure*}

We assume that the foreground model is used to subtract `point' sources with S~$>$~30~mJy, and estimate the amplitude of the power due to a fraction, $\xi$, of doubles being mis-subtracted, compared with the amplitude of the power when they are correctly subtracted. We estimate the power due to mis-subtracted doubles with flux densities, ${\rm S_{min}}=30{\rm ~mJy} < {\rm S} < {\rm S_{max}}=1$~Jy for $\xi$=0.11 and the Rayleigh distribution of doubles from Figure \ref{fig:GMRTseps}. Figure \ref{fig:residual_power} displays the ratio of the residual power to the total point source power ($P_{\rm PNT}=\mathcal{F}^\dagger {\bf C}_{\rm PNT} \mathcal{F}$; Equations \ref{cov_nunu}--\ref{eqn:fourier_prop}), and the residual power difference.
The power bias increases toward longer baselines, but is dominated by the foreground power in the wedge.
In Figure \ref{fig:residual_power_data} we show the equivalent ratio and difference plots for the 191 observations with the different source models peeled.
As expected, the impact is larger for the longer baselines, and the overall amplitude of the residual power is $\sim{10}^{-3}-{10}^{-4}$ for most of the parameter space. Unlike the predictions, which are noise-free and simple, the maximum relative difference is found in the EoR Window, where the use of the improved sky model during calibration and subtraction has reduced the power leakage from the wedge. Point source subtraction therefore has the potential to bias a cosmological EoR signal, particularly in the crucial EoR Window.

There are key similarities and differences between the model prediction in Figure \ref{fig:residual_power} and the data in Figure \ref{fig:residual_power_data}. Firstly, note that the data include radiometric noise, and therefore in regions where there is a balance of noise and foreground signal (primarily the EoR Window between $k_\parallel$=0.09--0.4 and above the wedge), we expect and find that the ratio is closer to unity. The copies of the foreground wedge at $k_\parallel$=0.45 (and above) are due to regular missing channels in the MWA bandpass and can be ignored. The key comparison is in the wedge, where we are foreground dominated. Here, the data show ratios of $<10^{-5}$ at small $k_\perp$, increasing to $<10^{-3}$ at large $k_\perp$, compared with the prediction of $<10^{-7}-10^{-3}$ over the same range. The lack of clear $k_\parallel$ dependence in Figure \ref{fig:residual_power} stems from it being a ratio, and therefore does not reflect the smaller numbers in the numerator and denominator outside of the wedge (compare with the RHS of Figure \ref{fig:residual_power}, which is a combination of the foreground wedge and the $k_\perp^4$ dependence). The key conclusion of this work is that the simple residual power model broadly reproduces the features observed in the data, and that longer baselines are much more heavily affected than short.

\begin{figure}
\begin{subfigure}[b]{1.1\columnwidth}
\includegraphics[width=0.95\columnwidth]{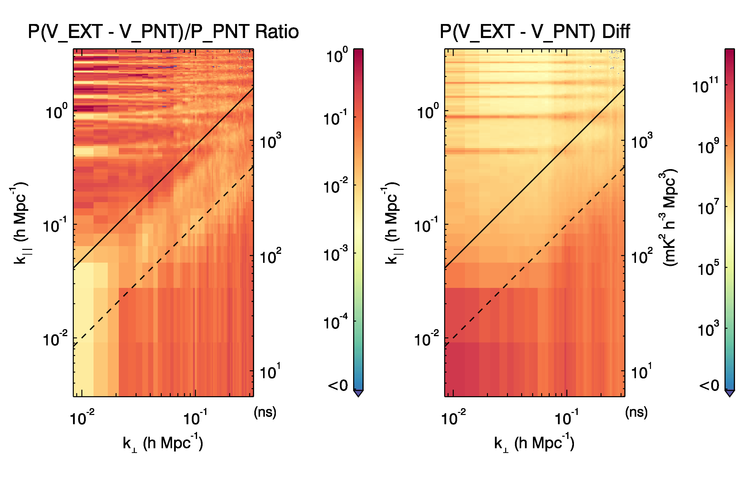}
\caption{}
\end{subfigure}
\begin{subfigure}[b]{1.1\columnwidth}
\includegraphics[width=0.95\columnwidth]{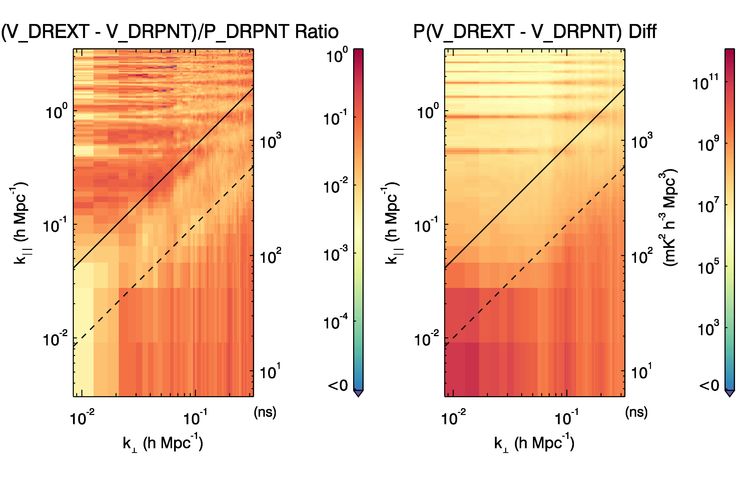}
\caption{}
\end{subfigure}
\begin{subfigure}[b]{1.1\columnwidth}
\includegraphics[width=0.95\columnwidth]{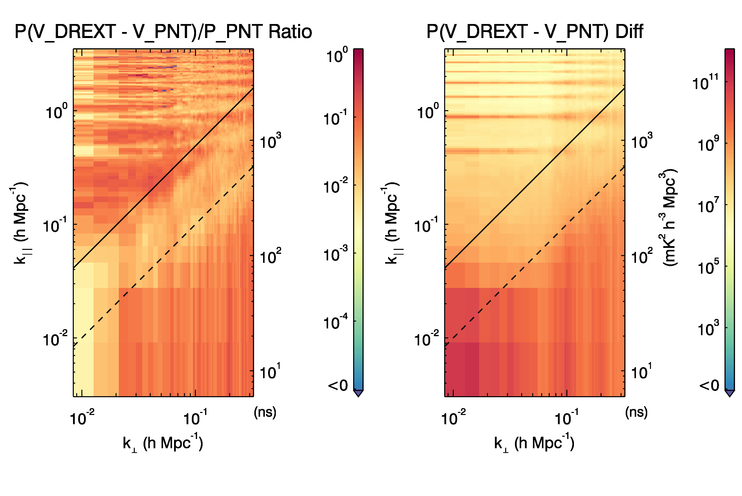}
\caption{}
\end{subfigure}
\caption{The ratios and differences for (a) Subtracting extended sources as extended versus point; (b) Subtracting double and extended sources as double and extended versus double and point; (c) Subtracting double and extended sources as double and extended versus point.}
\label{fig:full_comparison}
\end{figure}

A similar analysis can be performed for the extended source model. Figure \ref{fig:full_comparison} displays the ratios and differences for (a) Subtracting extended sources as extended versus point; (b) Subtracting double and extended sources as double and extended versus double and point; (c) Subtracting double and extended sources as double and extended versus point. The latter (c) shows the case where the full extended source model has been applied, compared with the standard full point source model, and therefore represents the best improvement. In all cases, a more correct subtraction model yields improved power removal, and the use of an extended model has significant impact on the results (compared with just using a model with double sources, where the impact is smaller).
Most notable is the impact in the EoR Window, where use of a model with extended sources removes a large amount of leaked power compared with treating the sources as simple point sources or doubles. 

It is important to point out that the EoR1 field is characterised by a large number of bright extended sources and those constitute the majority of the ten brightest sources. The combination of the extended morphology with their high surface brightness exacerbates the problem of their subtraction, stressing more the need for detailed models for these sources.

\section{Effect of improved model}
\label{sec:imp_mod}
\subsection{Data Processing}

Being originally designed to be the backend of the MWA, we chose to use the RTS to process the EoR1 field data. More importantly, the flexibility and the capabilities the RTS possesses for ingesting and handling a sky model with mixed formats make this software the perfect candidate to assess the improvements over the starting sky model.
 
The RTS makes use of direction-dependent beam response, ionospheric modelling and correction, and in-field calibration using several sources simultaneously. For each catalogue produced, we run the RTS twice: first we compute an optimal calibration solution for each pointing; then we perform the source subtraction. During calibration, the considered observation is used to compute the direction independent Jones matrices of each MWA tile. This is achieved fitting the uncalibrated visibilities with a model formed by the 500 (apparent, as they are attenuated by the primary beam) brightest sources in the field of view. 
At this stage of the processing these sources are combined in a single calibrator in order to achieve a high signal-to-noise ratio. For the same reason, we process the whole observation in one single time cadence, meaning that we integrate over time for the full duration of the observation.  

The second RTS run involves the subtraction of the sources through \textit{peeling} \citep{2004SPIE.5489..817N}. The advantage of this technique over more common source subtraction methods is that the sources are removed in the visibility instead of image space. This allows the possibility of performing further calibration against the considered source, with the direct consequence of improving its subtraction. Further, any improvement gained during this processing is reflected in the image space, where fewer artefacts due to non-optimal calibration may appear.

Although the RTS is capable of performing full peeling, due to the large number of sources to subtract the inclusion of a full calibration loop is not viable in our analysis. In fact, full peeling is only carried out on the brightest and most complex source in the field, Fornax A. All the other sources are treated in a slightly different way. While they share most of the processing steps used during full peeling (e.g. rotation of the visibilities to phase centre for the considered source, suppression of non-centred sources), the antenna-based gain calibration solutions are not computed for these calibrators. Instead, two phase gradient parameters (plus amplitude estimation) are used to model the ionosphere-induced phase ramp across the array for the source in question. This translates to a drastic reduction of parameters to be fitted in the model, hence the possibility to generate independent solutions for many more calibrators. Further, this method allows peeling of sources with low signal-to-noise ratio, whilst the same is not possible if performing full peeling. For simplicity, we will keep using the word `peeling' when referring to any source subtraction performed here.


During this step, we subtract the same 500 sources we combined in the calibrator model, but we do not perform any clustering so as to take full advantage of the RTS ionospheric correction on the single sources. 

\subsection{Results}

We first compare the calibration solutions obtained from each of the sky models. Although $\sim 20 \%$ of the sources in the calibrator cluster were replaced with high resolution models, due to the nature of the technique used during calibration we do not expect a difference between the various solutions.  
No appreciable difference in the slopes or dispersion of the gain phase values was found, and we obtained a slight reduction in the dispersion of the gain amplitude values when calibrating using the extended source model. This suggests that the performance of our calibration solution cannot be improved using this technique and adds confidence about the robustness of this processing step. 
It should be noted that the second strongest source in the field, PMN J0351-2744, is known to be polarised \citep{2013ApJ...771..105B} and that significant diffuse polarisation has been detected in the MWA EoR0 field \citep{2016ApJ...830...38L}. Although a polarisation analysis lies outside the scope of this paper, future works may make use of similar calibration techniques to rate the polarimetric behaviour and to probe the EoR1 field for diffuse polarisation.
A further quantitative evaluation can be carried out comparing source-free regions between the images obtained from the two data reduction processes. We select ten regions at different distances from the beam centre and compare the noise reading of each pair. In every case, we find a lower background noise characterising the image obtained with the extended models, with decrements ranging from $1$ to $8\%$ from the point source model background levels.

\begin{figure}
\begin{subfigure}[b]{1\columnwidth}
\includegraphics[width=\columnwidth]{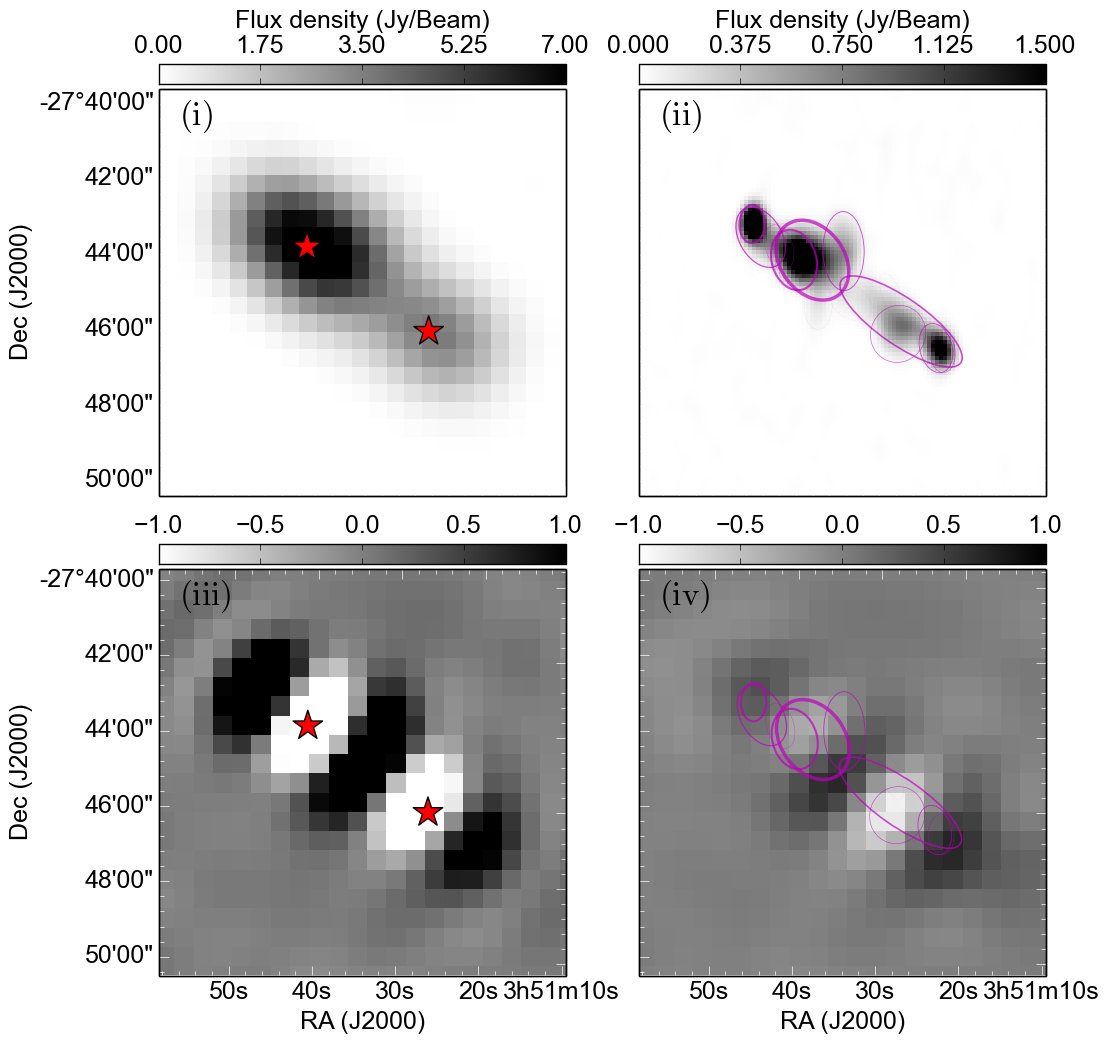}
\caption{PMN J0351-2744}
\end{subfigure}
\begin{subfigure}[b]{1\columnwidth}
\includegraphics[width=\columnwidth]{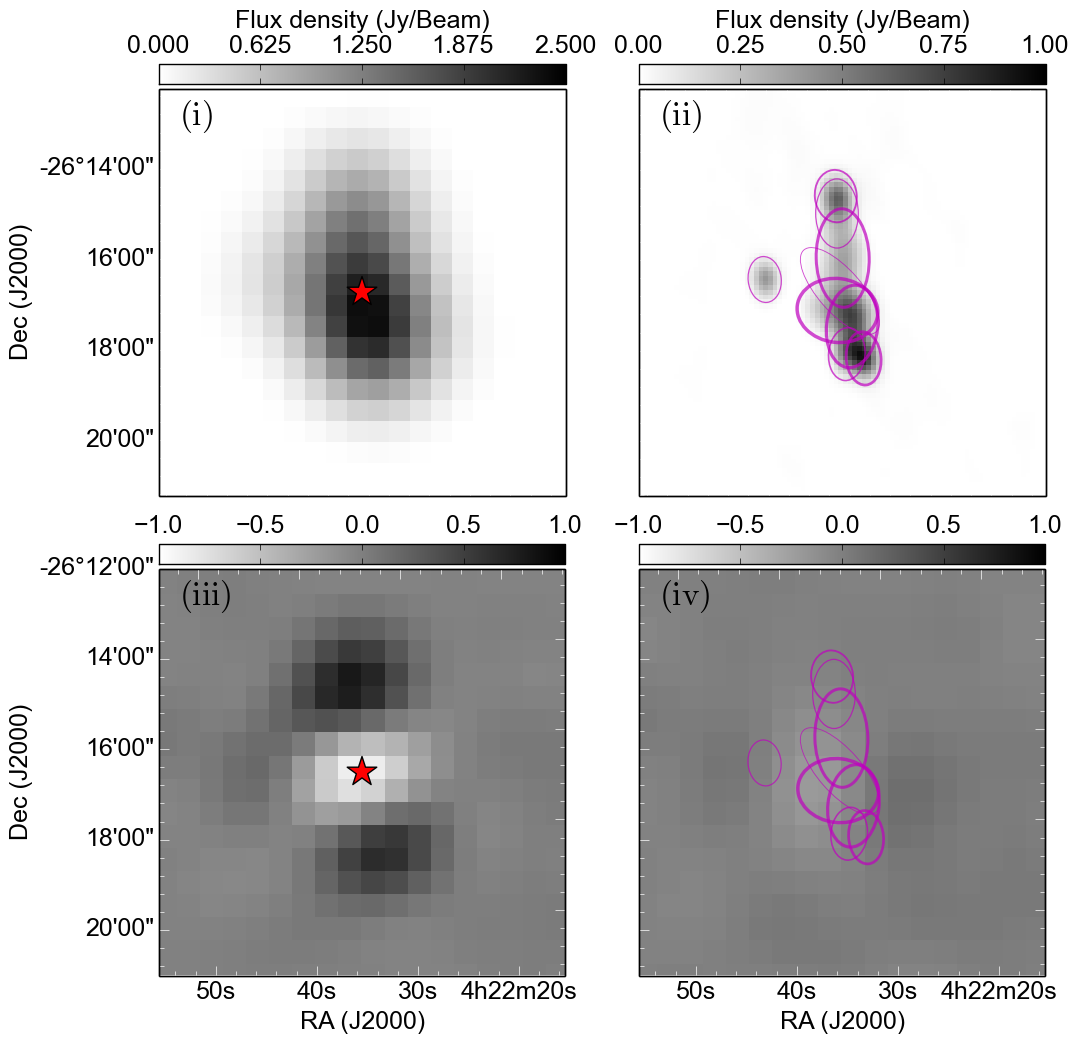}
\caption{PKS 0420-26}
\end{subfigure}
\caption{An example of the improvements obtained with the new models for the sources (a) PMN J0351-2744 (c.f Figure~\ref{fig:res_comp}) and (b) PKS 0420-26. In both figures: (i) GLEAM source positions are plotted on MWA data; (ii) \texttt{PyBDSM} Gaussian fits plotted over TGSS ADR1 data; (iii) the residuals left in MWA data after subtracting the  point source model; (iv) the residuals left in MWA data after subtracting the \texttt{PyBDSM} Gaussian extended model. In (ii) and (iv), the linewidths used to plot the Gaussian fits are scaled to the flux density of the Gaussian component for clarity. Note that PMN J0351-2744 is a $\sim28\,$Jy source before peeling.}
\label{fig:res1}
\end{figure}

On the other hand, we expect to find striking differences when comparing the residuals of the peeled extended model vs the point source model for the same source. For most of the sources, the Gaussian model resulted in a substantial improvement, already visible by eye when looking at the residuals after peeling the catalogues (e.g. Figure~\ref{fig:res1}). More quantitatively, we select regions covering the subtraction residuals and compute min, max, and root mean square (r.m.s.) of the pixel values (Table~\ref{table:residual_stats} shows these statistics for the five strongest sources). Being the selected area free of spurious sources, we assume that smaller r.m.s. values indicate a better subtraction of the source. 
As expected, the magnitude of the improvements varies from source to source and the most important weighting factor here is the morphology of the considered source. In fact, Gaussian modelling benefits the subtraction of the most extended sources the most and has the largest impact as residual r.m.s. near these sources.

\begin{figure}
\includegraphics[width=\columnwidth,scale=1.2]{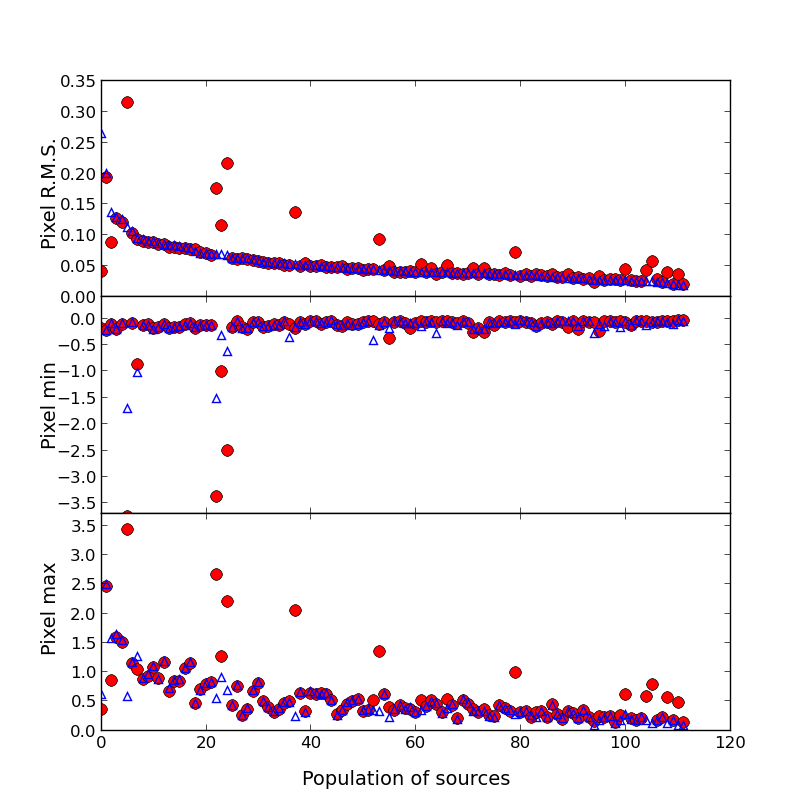}
\caption{Top panel: the r.m.s. of a region of $50\times50$ pixels over the source residuals is shown for each point source model (filled circles) and for the Gaussian models (empty triangles). Middle - bottom panel: the minimum and maximum pixel value respectively of the subtraction residual is plotted.  
In all the panels, the population of sources has been ordered with respect to the r.m.s. values of the Gaussian model residuals, in decreasing order.}
\label{fig:stats}
\end{figure}

We find that $\sim60\%$ of the residuals shows improvements when the extended models are used during the source subtraction. However, if we set the background noise level to $14$ mJy (average of four source-free regions near the beam centre of the primary beam corrected image) only 16 sources show a clear improvement, while the r.m.s. difference of the remaining 51 sources falls within the noise level\footnote{These numbers are retrieved using the r.m.s. of the residuals as a reference. Replicating the same computation using the min (or max) as a proxy for the differences the number of improved models rises sensibly.}. 
We find that apart from three cases, the same happens for the sources that seem to not benefit from the extended models. 
Figure~\ref{fig:stats} gives a summary of the results discussed above. 

\newcolumntype{P}[1]{>{\centering\arraybackslash}p{#1}}
\begin{table}
\renewcommand{\arraystretch}{1.2}
\centering
\caption{Pixel value statistics for five strong source residuals. The last column shows the r.m.s. computed on a source-free region around the considered one. For each region, the first row shows results when the data are processed using the catalogue with point source model, while the second row shows the same quantity for the extended model catalogue.}
	\begin{tabular}{P{1cm} P{1cm}P{1cm}P{1cm}P{1cm}P{1cm}}
	\hline
	\textbf{Region} & \textbf{min} & \textbf{max} & \textbf{rms} & \textbf{rms}  \\
	\hline \hline
	1 &  -2.447 & 2.151 & 0.522 & 0.032\\
	 & -0.608 & 0.664 &  0.151 & 0.029\\
	2 & -2.542 & 2.313 & 0.588 & 0.021\\
	 & -1.158 & 0.391 & 0.203 & 0.019\\
	3 & -0.388 & 0.750 & 0.206 & 0.028 \\
	 & -0.285 & 0.697 & 0.186 & 0.022\\
	4 & -0.582 & 0.727 & 0.199 & 0.016\\
	 & -0.188 & 0.103 & 0.053 & 0.015\\
	5 & -0.232 & 0.430 & 0.093 & 0.018 \\
	 & -0.059 & 0.188 & 0.051 & 0.016\\
	\hline \hline
	\end{tabular}
\label{table:residual_stats}
\end{table}

\subsection{Impact on EoR power spectrum}\label{sec:ps_results}

Ultimately, we want to assess if precise source modelling has an impact on EoR cosmological signal detection. As briefly discussed in Sec. \ref{sec:mis-sub}, the analysis in $k$-space predicts that extended sources are able to push more power toward larger $k$ modes compared with the case in which only point sources are present in the model and in practice contribute more power on all scales in the PS. Overall we find slightly more power left over at larger $k_\bot$ modes when subtracting the sources incorrectly. The results show qualitative agreement with the theoretical predictions based on the same dataset, although big differences start to occur at $k_\bot$ modes slightly higher than we are sampling. 

When comparing the processing that includes both extended and double sources, with the processing applying a point source model alone, the power improvement in the EoR Window ($0.01<k_\bot{<}0.1$~Mpc$^{-1}$, $0.08<k_\parallel{<}0.4$~Mpc$^{-1}$) is $\sim{2}\times{10}^8$~mK$^2$~Mpc$^3$. This represents a factor of $\sim$2 improvement. However, the thermal noise level in these data is $\sim{1}\times{10}^7$~mK$^2$~Mpc$^3$, and therefore residual foreground contamination occurs in the improved model. It is difficult to assess the impact of further improvement of the foreground model.

We can pose an additional question: which of the extended sources are contributing most to the improvement? The brightest ten sources that are modelled differently in each calibration and peeling test, have flux densities of 4.6--26.6~Jy, whereas the full set different sources have flux densities extending down to $\sim$30~mJy ($\sim$500 sources). The analytic model, using a realistic source number density model for bright and weak sources, predicts that 90\% of the improvement observed in the data (the factor of two power improvement), can be attributed to these ten brightest sources. Therefore, as one may intuitively expect, it is the brightest sources being mis-modelled that are the most important for the additional foreground power, and therefore careful models for these are crucial.

At this point, it should be noted that the residuals left by Fornax A are comparable, in terms of flux density, to those left by the brightest extended sources after applying the improved models. The pixel peak values of the brightest areas in the residuals are typically below 0.5~Jy, whilst most of the source is removed down to the noise level. Although these residuals still contribute to foreground power in the PS, we kept this model unchanged in all our sky models. Hence, while a deeper analysis could show the limits of the current model, this is not affecting the \textit{relative} improvements between the sky models and thus the overall results of this work.

In conclusion, our analysis suggests that detailed models are needed for extended and resolved double sources when those have to be subtracted. Subtracting them as point sources leaves residual excess power and therefore has the potential to bias the signal. Although this may be a strategy worth applying, we find that it is more important to use detailed models for the extended brightest sources, and that subtracting doubles below the confusion noise using point source models has a lesser impact on the final result.

\section{Discussion}
\label{sec:disc}

Ultimately, we note that the MWA uses only visibilities coming from the core antennas for the EoR experiment and all of them are less than 2\,km length. The results obtained in this paper suggest that once we used a detailed sky model, the quality of the processed data improved even though our dataset is intrinsically limited in resolution due to the baseline cut-off.
The detailed sky model was derived from TGSS data, but in a similar fashion we could have taken advantage of the full MWA array and use the long baselines to improve the models. It is possible to relate the exercise carried out in this paper with other experiments that deliberately do not include long baselines in their design, e.g. PAPER \citep{parsons10}  and HERA \citep{deboer16}. The improvement that the longer baselines bring to the EoR MWA sub-array in terms of foregrounds modelling and calibration can be mimicked by building outriggers to get longer baselines and so to obtain a better PSF. Those ideas are discussed in details in \citet{dillon16}. 



The same thought can be elaborated in light of the future realisation of the Square Kilometre Array (SKA). We argue that the realisation of a detailed sky model for an SKA-Low EoR experiment would allow for the exclusion of the long baselines for the configuration of a possible sub-array. This would cause a dramatic reduction in the computational power required for the data reduction and an overall simplification of the end-to-end processing.
The analysis presented here can be applied to an SKA model, where the residual power in the EoR window from mis-subtracting double sources can be estimated as a function of maximum baseline (array resolution). While increasing the maximum baseline reduces the characteristic separation of doubles ($\sigma$), it also decreases the fraction of non-point sources that will not be measurable. This is most pronounced at 50~MHz, where the thermal noise and confusion limit are high, and angular resolution is poorest. It is at this lowest frequency where we estimate the angular resolution required such that mis-modelled doubles with smaller separations contribute less power bias than the thermal noise level for the EoR/CD power spectrum experiment (assuming a 1000~h observation).

Equation \ref{cov_nunu} shows that the power is proportional to $\xi{k_\perp}^4\sigma^4$. Increasing the maximum baseline of the array ($B$) improves the spatial resolution, and this consequently shifts the peak of the distribution of source separations (Figure \ref{fig:SIdist}) with $\sigma\propto 1/B$. We compute the value of $B$ for which the additional power bias from remaining unresolved double sources is less than the thermal noise expectations for a 1000~h SKA experiment at 50~MHz \citep[1~mK$^2$ at $k_\bot=0.1$~Mpc$^{-1}$, Fig 3.,][]{koopmans15}. I.e., $P_{\rm res} = 1$~mK$^2$. We note that high signal-to-noise sources can be better-resolved, and therefore choose 10$\times$ the confusion limit to set the maximum source flux density to be considered \citep[1~mJy/beam, Fig 16.,][]{braun_error}. We find that a maximum baseline of 35--50~km is sufficient such that double sources can be resolved by the instrument and included in a global sky model. We therefore predict that the current 45--65~km maximum baseline model for SKA1-Low will be sufficient, even at the most challenging lowest frequencies.

\section{Conclusions}
\label{sec:conc}

In this paper, we performed a complete analysis of the effects obtained by using precise foreground models, in the context of EoR experiments. 
We focused our analysis on a region of the sky, the MWA EoR1 field, observed for the search of the HI cosmological signal. 

We identify sources in the field that have multiple counterparts in higher-frequency radio catalogues and used high-resolution data from the TGSS survey at 150~MHz to create improved models for this population of sources. We find that 11\% of the total number of matches show as double sources in the high-resolution data and that 114 sources show a more complex morphology, needing a more detailed modelling. 
Then, building the detailed sky model through multiple iterations, we evaluate the improvements over a more simplistic one based on point sources.
Using the high-resolution data, we find an improvement in calibration accuracy of up to 8\% and an overall reduction in the residuals after subtraction by up to a factor of four.

The power spectrum analysis shows improvements aligned with those found in image space. First, we evaluate how the effect of mis-subtracting non-point sources as point sources and subtracting them properly propagates in the power spectra. As Figure~\ref{fig:residual_power_data} shows, we find that the point source subtraction has the potential to bias the cosmological EoR signal, because the use of the correct models for the double sources reduces the power leakage from the wedge, where foregrounds dominate. Then, we perform a similar analysis to the extended source model. Panel (c) of Figure~\ref{fig:full_comparison} shows the comparison between the use of the point source model and the one including both extended and double sources. 

We find an improvement of $\sim{2}\times{10}^8$~mK$^2$~$h$Mpc$^3$ in the EoR window when using the high-resolution data, which corresponds to an improvement of a factor of $\sim2$. We also find that, using a realistic source number density model for bright and faint sources, the analytic model predicts that $\sim90$\% of that improvement can be attributed to the ten brightest sources, which have flux densities that range from 4.6 to 26.6~Jy.


In conclusion, we state that extended or multi-component sources that are above the confusion noise need a detailed model for their subtraction, for subtracting them as point sources removes an excess of power that biases the signal in several k modes.

In this analysis we have compared the use of point source models with multi-component models based on higher resolution observations.  However we have not explored the use of extended source models based on lower resolution data such as is already included in the GLEAM catalogue.  This important future study is needed to better address the question of whether longer baselines are needed in telescopes designed to detect the EoR signal.

\appendix
\section{Appendix}
Here we derive the error in the visibility variance due to subtracting double sources as point sources, considering the doubles to be randomly-oriented in angle on the sky, and have a Rayleigh-distribution of separations, with a known characteristic scale, $\sigma$. From Equation \ref{eqn:probs}, we have:
\begin{eqnarray}
&&{\rm var}(V(u,v)) \propto  \\\nonumber
&&\int_{2\pi} p(\phi)d\phi \int_{\Delta{r}} p(\Delta{r})d\Delta{r} \, (\cos{\Delta{r}\pi(u\cos{\phi} + v\sin{\phi})}-1)^2,
\end{eqnarray}
where the angle, $\phi$, and separation, $\Delta{r}$ distributions are:
\begin{eqnarray}
\phi &\sim& \mathcal{U}(0,2\pi]\\
\Delta{r} &\sim& \mathcal{R}(r;\sigma).
\end{eqnarray}
For sufficiently small $\Delta{r}$ ($\ll\!\sqrt{u^2+v^2}$), met for baselines shorter than $\sim$~5~km, we expand the cosine term to separate the $u$ and $v$ term, and Taylor-expand to quadratic order;
\begin{eqnarray}
&\cos{(a+b)} = \cos{a}\cos{b} - \sin{a}\sin{b}\\
&= \cos{(\pi\Delta{r}u\cos\phi)}\cos{(\pi\Delta{r}v\sin\phi)} - \\\nonumber &\sin{(\pi\Delta{r}u\cos\phi)}\sin{(\pi\Delta{r}v\sin\phi)}\\
&\simeq 1 - \frac{(\pi\Delta{r}u\cos\phi)^2}{2} - \frac{(\pi\Delta{r}v\sin\phi)^2}{2}\\\nonumber
&- (\pi\Delta{r}u\cos\phi)(\pi\Delta{r}v\sin\phi),\\
&= 1 - \frac{\pi^2\Delta{r}^2}{2}\left(u^2\cos{\phi}^2 +v^2\sin{\phi^2}\right),
\end{eqnarray}
such that;
\begin{equation}
\left(\cos{\Delta{r}\pi(u\cos{\phi} + v\sin{\phi})}-1\right)^2 \simeq \frac{\Delta{r}^4\pi^4}{4}(u^2\cos{\phi}^2 + v^2\sin{\phi}^2)^2.
\end{equation}
Expanding the quadratic in parentheses, using the power-reduction formulae for trigonometric functions, and performing the integrals over angle $\phi$, we find:
\begin{eqnarray}
&&{\rm var}(V(u,v)) \propto \\\nonumber
&&\int_0^{\infty} d\Delta{r} \frac{3\pi^5\Delta{r}^5(u^2+v^2)^2}{8\sigma^2} e^{\left( \frac{-\Delta{r}^2}{2\sigma^2} \right)}\\
&=& \frac{3\pi^5}{8}(u^2+v^2)^2\sigma^4.
\end{eqnarray}

We are considering a continuum source population, and so need to extend this expression to include the flux density contribution from these sources, and to study the covariance between different frequency channels (the primary dimension over which the contributions from a given source are correlated). Following \citet{trott16}, the covariance between different frequency channels at a given scale $u$ relates the different primary beam responses ($B(l;\nu)$) and the different Fourier kernels at each frequency, such that\footnote{We note that the residual visibility term, and integration over the beam term, decouple because the residual term is computed for point sources, which are assumed to be statistically uncorrelated between any two points in the field. They are also uncorrelated between different flux density bins. The only correlation of importance is that across frequency (because they are continuum sources), and this is the reason that the data covariance matrix, ${\bf C}_{\rm res}$, is computed spectrally.}:
\begin{eqnarray}
&&{\bf C}_{\rm res}(u,v;\nu,\nu^\prime) = \frac{3\pi^5}{8}(u^2+v^2)^2\sigma^4 \int_{S_{\rm min}}^{S_{\rm max}} S^2 \frac{dN}{dS} dS \nonumber\\
&& \times \displaystyle\iint_{d\Omega} d{\bf l} \, B({\bf l};\nu)B({\bf l};\nu^\prime) \, e^{\left( \frac{-2{\pi}i}{\nu_0}\Delta\nu({\bf u}\cdot{\bf l}) \right)} d\Omega.
\end{eqnarray}
The key components of this expression are: (1) the baseline-length dependence of the residual visibilities; (2) the source flux density contribution, summing the contribution to the variance of sources in each flux density bin between some lower and upper flux density limit; (3) the correlation between baselines formed from a given antenna pair at different frequencies (where the frequency changes the dimensionless baseline length, $u$), integrated over the full field of view, $d\Omega$ (sr).

The flux density limits denote the upper (brightest source in the field) and lower (limit of peeling of sources inaccurately) limits contributing to the residual. Performing the integration over flux density yields:
\begin{eqnarray}
&&{\bf C}_{\rm res}(u,v;\nu,\nu^\prime) = \frac{\alpha}{3-\beta}\frac{S_{\rm max}^{3-\beta}-S_{\rm min}^{3-\beta}}{S_0^{-\beta}}\left( \frac{3\pi^5}{8}(u^2+v^2)^2\sigma^4 \right) \nonumber\\
&& \times \displaystyle\iint d{\bf l} \, B({\bf l};\nu)B({\bf l};\nu^\prime) \, e^{\left( \frac{-2{\pi}i}{\nu_0}\Delta\nu({\bf u}\cdot{\bf l}) \right)}.
\end{eqnarray}
Finally, we include a simple, frequency-dependent Gaussian-shaped primary beam with characteristic width, $A(\nu)=(c\epsilon)/(\nu{D})$ ($\epsilon\simeq{0.42}$ converts the Airy disk FWHM to the $\sigma$ of a Gaussian), yielding:
\begin{eqnarray}
&&{\bf C}_{\rm res}(u,v;\nu,\nu^\prime) = \frac{\alpha\xi}{3-\beta}\frac{(S_{\rm max}^{3-\beta}-S_{\rm min}^{3-\beta})}{S_0^{-\beta}}\frac{\pi{c^2}\epsilon^2}{D^2}\\\nonumber
&&\times \left(3\pi^5(u^2+v^2)^2\sigma^4 \right) \frac{1}{8(\nu^2 + \nu^{\prime{2}})} e^{\left( \frac{-u^2c^2f(\nu)^2\epsilon^2}{4(\nu^2 + \nu^{\prime{2}})D^2} \right)},
\end{eqnarray}
where $f(\nu) = (\nu-\nu^\prime)/\nu_0$ captures the difference in frequency channels.

\begin{acknowledgements}


CMT is supported under the Australian Research Council's Discovery Early Career Researcher funding scheme (project number DE140100316). The Centre for All-sky Astrophysics (an Australian Research Council Centre of Excellence funded by grant CE110001020) supported this work.
This scientific work makes use of the Murchison Radio-astronomy Observatory, operated by CSIRO. We acknowledge the Wajarri Yamatji people as the traditional owners of the Observatory site. Support for the operation of the MWA is provided by the Australian Government (NCRIS), under a contract to Curtin University administered by Astronomy Australia Limited. We acknowledge the Pawsey Supercomputing Centre which is supported by the Western Australian and Australian Governments. We thank the staff of the GMRT that made these observations possible. GMRT is run by the National Centre for Radio Astrophysics of the Tata Institute of Fundamental Research. We acknowledge the International Centre for Radio Astronomy Research (ICRAR), a Joint Venture of Curtin University and The University of Western Australia, funded by the Western Australian State government.

\end{acknowledgements}

\newcommand{\pasa}{PASA}
\newcommand{\apj}{ApJ}
\newcommand{\aj}{AJ}
\newcommand{\apjl}{ApJ}
\newcommand{\aap}{A\&A}
\newcommand{\mnras}{MNRAS}
\newcommand{\aaps}{A\&AS}
\newcommand{\pasp}{PASP}
\newcommand{\physrep}{PhR}
\newcommand{\prd}{PhRvD}
\newcommand{\araa}{ARA\&A}
\newcommand{\procspie}{Proc. SPIE}
\bibliographystyle{apj}
\bibliography{ms_final}
\end{document}